\documentclass[twocolumn]{aastex631}

\usepackage{hyperref}
\usepackage{amsmath}
\usepackage{soul}
\usepackage{enumitem}

\begin{document}

\title{Magnetic Field Evolution of the Solar Active Region 13664}

\author{Robert Jarolim}
\affiliation{High Altitude Observatory, NSF NCAR, 3080 Center Green Dr., Boulder, CO 80301, USA}

\author{Astrid M. Veronig}
\affiliation{University of Graz, Institute of Physics, Universitätsplatz 5, 8010 Graz, Austria}
\affiliation{University of Graz, Kanzelhöhe Observatory for Solar and Environmental Research, Kanzelhöhe 19, 9521 Treffen am Ossiacher See, Austria}

\author{Stefan Purkhart}
\affiliation{University of Graz, Institute of Physics, Universitätsplatz 5, 8010 Graz, Austria}

\author{Peijin Zhang}
\affiliation{New Jersey Institute of Technology, 323 Dr Martin Luther King Jr Blvd, Newark, NJ 07102, USA}

\author{Matthias Rempel}
\affiliation{High Altitude Observatory, NSF NCAR, 3080 Center Green Dr., Boulder, CO 80301, USA}



\begin{abstract}

On 2024 May 10/11, the strongest geomagnetic storm since November 2003 has occurred, with a peak Dst index of $-412$ nT. The storm was caused by NOAA Active Region (AR) 13664, which was the source of a large number of coronal mass ejections and flares, including 12 X-class flares. Starting from about May 7, AR 13664 showed a steep increase in its size and (free) magnetic energy, along with increased flare activity. In this study, we perform 3D magnetic field extrapolations with the NF2 nonlinear-force free code based on physics informed neural networks \citep{jarolim2023nf2}. In addition, we introduce the computation of the vector potential to achieve divergence-free solutions. We extrapolate vector magnetograms from SDO/HMI at the full 12 minute cadence from 2024 May 5-00:00 to 11-04:36 UT, in order to understand the active regions magnetic evolution and the large eruptions it produced. The computed change in magnetic energy and free magnetic energy shows a clear correspondence to the flaring activity. Regions of free magnetic energy and depleted magnetic energy indicate the flare origin and are in good correspondence with observations in Extreme Ultraviolet. Our results suggest that the modeled solar flares are related to significant topological reconfigurations. We provide a detailed analysis of the X4.0-class flare on May 10, where we show that the interaction between separated magnetic domains is directly linked to major flaring events. With this study, we provide a comprehensive data set of the magnetic evolution of AR 13664 and make it publicly available for further analysis.

\end{abstract}



\section{Introduction} 
\label{sec:intro}

Solar flares and coronal mass ejections (CMEs) are the most energetic phenomena in our solar system, and can cause severe effects on the space weather at Earth and other solar-system planets. 
Flares and CMEs are different facets of the same physical process. They are known to result from instabilities in the coronal magnetic field and the impulsive release of vast amounts of energy by magnetic reconnection \citep{priest2002,Schrijver2009}, which is subsequently converted into kinetic energy of high-energy particles, plasma motions and heating \cite[e.g.,][]{Veronig2005,Fletcher2011}.
The magnetic energy that is suddenly released in solar flares (on time scales of minutes to hours) has been  previously accumulated and stored (on time scales of days to weeks), through the emergence of magnetic flux from the convection zone to the solar atmosphere and by shearing motions producing strong electric currents in the Active Region's (AR) corona \citep[e.g.,][]{forbes2006,sun2012ar11158,wiegelmann2014}. Therefore, it is important to model the coronal magnetic field and to study the 3D topology of ARs that may lead to major flaring \citep[][]{wiegelmann2014,Janvier2015, Korsos2024preeruption}.

Statistical studies have shown that large flares are preferentially produced by large ARs of high magnetic complexity \citep[e.g.,][]{Leka2007,schrijver2007}. \cite{sammis2000} found that about 60\% of the X-class flares (100\% of flares $\ge$X4) resulted from ARs of Mount Wilson magnetic class $\beta \gamma \delta$ and a size $>$1000~$\mu$hem. 
However, where the bulk of the magnetic energy is stored, and where and how the energy release in flares is triggered is still a topic of intense research \cite[e.g.,][]{green2018,kusano2020flares,gupta2021energy_budget}.

During 2024 May 2 to 14, NOAA AR 13664 was visible from Earth, and developed to one of the largest and most flare-productive ARs in the recent decades. From May 4 to 7, it grew in size from about 110 to 2700 $\mu$hem \cite[see the overview in][]{Hayakawa2024}, and starting from May 6 it was of  magnetic class $\beta\gamma\delta$. Over its lifetime, AR 13664 produced 12 X-class flare (including an X8.7 flare when it has already rotated behind the Western limb) and 52 M-class flares. Noteworthy, its high activity caused the strongest geomagnetic storm since November 2003, with a peak Dst index of $-412$~nT on 2024 May 11, around 4~UT. 

The exceptional flaring activity and the fast evolution of AR 13664 while it was on the Earth-facing hemisphere, makes it an ideal candidate for a detailed investigation of the evolution of its magnetic complexity, the energy storage in the AR and the energy release during the major flares it produced. 

In this study, we model the ARs magnetic field during its transition across the solar disk and analyze the magnetic topology and magnetic energy build-up/release mechanisms. In Sect.~\ref{sec:method}, we introduce the used data, magnetic model, and analysis methods. Section \ref{sec:results} summarizes the result of our magnetic field extrapolations and the connection to the flaring activity of AR 13664. We investigate the energy build-up and release processes and specifically focus on the X4.0 flare on 2024 May 10. Finally, we provide a summary and interpretation of the model results (Sect.~\ref{sec:discussion}).

\begin{figure*}
    \centering
    \includegraphics[width=\linewidth]{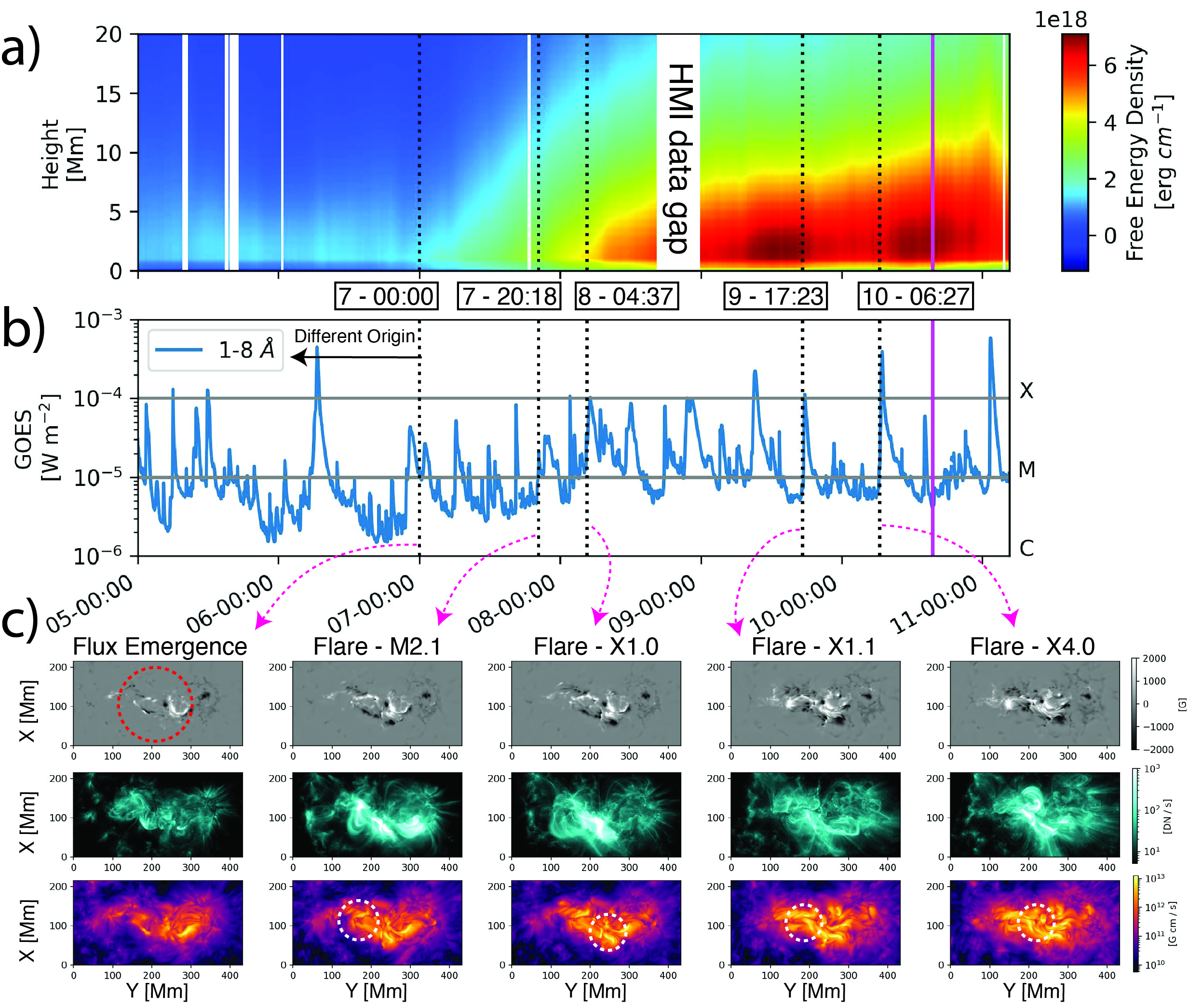}
    \caption{Overview of the evolution of NOAA AR 13664 from 2024 May 5-00:00 to 11-04:36 UT. a) Evolution of the vertical distribution of free magnetic energy. b) 
    GOES 1-8~{\AA} soft X-ray flux. c) Observed SDO/HMI radial magnetic field maps, SDO/AIA~131~{\AA} EUV maps, and the modeled  current density integrated along the vertical axis. We show the initial flux emergence (snapshot at 2024 May 7 00:00~UT) and the major flare eruptions (indicated by dashed lines). The magenta line indicates the point where the right edge of the SHARP reaches 60$^\circ$ longitude, beyond which the reliability of the vector magnetograms and the modeled coronal field decreases. The circles in the integrated current density maps indicate points of flux domain interactions. An animation of the full time series is available in the online journal. Left top: B$_z$ of the modeled photospheric magnetic field. Right top: EUV observations from SDO/AIA~131~{\AA} EUV. Left bottom: maps of integrated current density. Right bottom: maps of integrated free magnetic energy. The movie shows the re-configuration of magnetic domains (dark lines in current density maps) during strong solar eruptions.}
    \label{fig:overview}
\end{figure*}

\section{Method}
\label{sec:method}

In this study, we primarily examine the magnetic topology and evolution of AR 13664. Non-linear Force-Free (NLFF) extrapolations are frequently applied to obtain a realistic estimate of the coronal magnetic field from photospheric vector magnetograms \citep[e.g.,][]{wiegelmann2021sffmf, wheatland2011nlff, wiegelmann2010methood}. However, NLFF extrapolations typically require additional pre-processing of observational data and are computationally expensive \citep{wiegelmann2006preprocessing, wiegelmann2021sffmf}. 

Physics-Informed Neural Networks \citep[PINNs;][]{raissi2019pinns} are a novel method for solving partial differential equations and have the ability to smoothly integrate noisy data and incomplete physical models \citep{karniadakis2021physicsml}. 
In \cite{jarolim2023nf2}, PINNs were introduced for NLFF extrapolations, and demonstrated the ability to provide reliable magnetic field extrapolations in quasi real-time. This specifically enables the efficient computation of extrapolation series at a high temporal resolution.

A critical aspect of the NLFF methods is the divergence free condition, which is typically part of the optimization method \citep[e.g.,][]{wiegelmann2012parameters}. Here, we build on the PINN implementation from \cite{jarolim2023nf2} and introduce in addition extrapolations through the vector potential, instead of directly modeling the magnetic field (Sect. \ref{sec:method_nlff}). With this approach we can intrinsically obtain solenoidal magnetic field solutions.

\subsection{Data}
\label{sec:method_data}

We use vector magnetograms from the Helioseismic and Magnetic Imager \citep[HMI; ][]{schou2012hmi} onboard the Solar Dynamics Observatory \citep[SDO;][]{pesnell2012sdo}. Specifically, we use Space Weather Active Region Patches \citep[SHARP;][]{bobra2014sharps} available with a cadence of 12~min. Here, the vector magnetograms are reprojected to a Cylindrical Equal Area (CEA) grid with B$_x$, B$_y$, B$_x$, which we use as input for our magnetic field extrapolations. For our extrapolations we use the full resolution with a spatial scale of 0.36 Mm per pixel. Note that this resolution is typically not reproduced by our method, due to the disagreement between the force-free assumption and the observation. However, extrapolating the data at full resolution causes no significant increase in computing time. In addition, we use the provided error maps as input to our extrapolation method.

For the evaluation, we use observations in Extreme Ultraviolet (EUV) spectral bands from the Atmospheric Imaging Assembly \citep[AIA;][]{lemen2012aia} onboard SDO. To better visualize flaring activity, we create integrated EUV maps, where we use SDO/AIA observations from a single spectral band at a one minute cadence, and integrate the resulting image stack along the temporal axis. We specifically use the AIA 94 and 131~{\AA} filters, which are most sensitive to the hot flaring plasma, with peaks in the temperature response curves at $T \approx 7$ and $T \approx 11$ MK, respectively.

For our study, we use observations from 2024 May 5-00:00 to 11-04:36~UT. The series contains SDO/HMI data gaps. The only significant data gap occurs between 8-16:36 and 9-00:00, which coincides with several M-class flares and an X1.0 flare (8-21:08). Therefore, these events were not considered in our analysis.

\subsection{Non-linear force-free extrapolations}
\label{sec:method_nlff}

For the Non-Linear Force Free (NLFF) extrapolations of AR 13664, we use the method from \cite{jarolim2023nf2}, which is based on PINNs.
Here, a Neural Network is trained to act as a function approximation of the modeled magnetic field $\Vec{B}$. The model optimization is performed by iteratively sampling points from the boundary condition and randomly from within the simulation volume (x, y, z). The coordinate points are used as input to the neural network and mapped to the corresponding magnetic field vector (B$_x$, B$_y$, B$_z$). For the boundary condition the network is optimized to match the observed magnetic field vector, while for the randomly sampled points the residuals of the force-free equation
\begin{equation}
\label{eq:l_ff}
    L_{\rm ff} = \frac{\lVert(\vec{\nabla} \times \vec{B}) \times \vec{B}\rVert^2}{\lVert \vec{B} \rVert^2 + \epsilon},
\end{equation}
and the divergence-free equation
\begin{equation}
\label{eq:l_div}
    L_{\rm div} = |\vec{\nabla} \cdot \vec{B}|^2,
\end{equation}
are minimized. The derivatives are computed by automatic differentiation and used to construct the partial differential equations. Given that the neural network is fully-differentiable, automatic differentiation can be used to compute smooth derivatives of the model outputs with respect to the input coordinates.  As shown in \cite{jarolim2023nf2}, typically it is not possible to achieve a perfect agreement between the observations and the force-free model, and consequently a trade-off between the two conditions is found. This shortcoming is intrinsic to NLFF extrapolations, and can only be mitigated by building on a more complex physical model (e.g., magneto-hydrodynamic modeling) or by additional observational constrains \citep[e.g., chromospheric magnetic field measurements;][]{jarolim2024multi, fleishman2019, yelles2012chromospheric_extrapolation}.

For the optimization of the boundary condition, we take the provided error maps into account. Here, we compute the difference between the modeled ($\vec{B}$) and the reference magnetic field ($\vec{B}_{0}$)
\begin{equation}
    \vec{B}_{diff,0} = \text{abs}(\vec{B} - \vec{B}_{0}).
\end{equation}
We account for uncertainties in the measurement by subtracting the error map $\vec{B}_{error}$ and clipping negative values
\begin{equation}
    \vec{B}_{diff,clipped} = \max \{\vec{B}_{diff,0} - \vec{B}_{error}, 0\}.
\end{equation}
Therefore, we only optimize values where the modeled field exceeds the error threshold.
For the loss of the boundary magnetic field, we compute the vector norm of the clipped difference vector
\begin{equation}
\label{eq:l_b}
    L_{B} = \lVert \vec{B}_{diff,clipped} \rVert^2.
\end{equation}

In this study, we replace the direct modeling of the magnetic field $\Vec{B}$, with the computation of the vector potential $\Vec{A}$ which is derived by taking the curl of $\Vec{B}$
\begin{equation}
    \vec{B} = \nabla \times \vec{A}.
\end{equation}
From the fundamental relation of vector calculus this implies $\nabla \cdot B = \nabla \cdot (\nabla \times A) = 0$, which directly leads to solenoidal field solution. Note that this computation is performed with auto-differentiation and that a finite-differences approach will show deviations from a perfectly solenoidal field.

Omitting the optimization for the divergence-free condition, we obtain the final loss from the force-free and boundary condition
\begin{equation}
    L = \lambda_{\rm ff} L_{\rm ff} + \lambda_{\rm B} L_{\rm B}.
\end{equation}
For our initial extrapolation we exponentially decay $\lambda_{\rm B}$ from $1,000$ to $1$ over $50,000$ iterations. We throughout set $\lambda_{\rm ff}$ to 0.1.

\subsection{Metrics}

From the resulting magnetic field extrapolations we transfer the neural representation to a grid representation by sampling each point in our simulation volume at a resolution of 0.72 Mm per pixel, which corresponds to a rebinning by a factor of 2 from the original SDO/HMI resolution. In addition to the magnetic field vector $\Vec{B}$, we compute the current density $\Vec{J}$ according to Ampere's law
\begin{equation}
    \vec{J} = \frac{c}{4 \pi} \nabla \times \vec{B},
\end{equation}
where $c$ corresponds to the speed of light. Here, we use directly  automatic differentiation of the neural representation, to obtain smooth derivatives that are independent of the spatial grid resolution. Note that computed derivatives are in units of the normalized input coordinates and are converted to physical units (Mm) for our evaluation.

From the resulting cubes we compute measurements of magnetic energy, free magnetic energy, and the quality metrics.
The magnetic energy is defined as
\begin{equation}
    E = \sum_i^N \frac{B_i^2}{8\pi},
    \label{eq:energy}
\end{equation}
where $i$ refers to the $i$th grid cell and N to the total number of grid cells. The free magnetic energy corresponds to the difference between the magnetic energy of the NLFF field and the potential field
\begin{equation}
    E_{free} = E_{FF} - E_{PF}.
     \label{eq:free_energy}
\end{equation}
Here, we compute the potential field solution using the Green's function approach as proposed by \cite{sakurai1982green}. As input for the potential field extrapolation we use the modeled bottom boundary of our NLFF solution (i.e., the adapted boundary).

For the visualization of spatial distribution we compute the local magnetic energy density ($E_i = {B_i^2} / {8\pi} $), and integrate along the vertical or horizontal axis. We compute energy difference maps analogously, by calculating the energy difference per grid cell ($\Delta E_i = E_{\text{t2}, i} - E_{\text{t1}, i}$) and integration along the vertical axis. The total energy difference is computed by integrating over the full simulation domain
\begin{equation}
    \Delta E_{\text{total}} = \sum_i^N \Delta E_i \, .
    \label{eq:E_total}
\end{equation}

To provide an upper estimate of the released magnetic energy during flares, we compute the total depleted magnetic energy $\Delta E_{-}$ by clipping local energy increases prior to the integration:
\begin{equation}
    \Delta E_{-} = \sum_i^N \min(0, \Delta E_i) \, .
    \label{eq:E_negative}
\end{equation}
This excludes the energy build-up that may be continuously ongoing in the AR also during the flaring time (e.g., due to flux emergence). However, we note that also spatial redistribution may falsely count towards released magnetic energy (e.g., movement of magnetic elements). Therefore, this metric can only provide an upper estimate of released magnetic energy, while more advanced metrics would be needed to discern between the parallel processes.

We compute metrics for divergence- and force-freeness to verify the validity of our method. For this we use the normalized divergence
\begin{equation}
    L_{div,n}(\vec{B}) = \sum_i |\vec{\nabla} \cdot \vec{B}_{i}| / \lVert \vec{B}_{i} \rVert \,.
\end{equation}
which we compute from the sampled grid based on finite differences. In addition, we compute the current weighted angle between the magnetic field and the current density $\theta_{\text{J}} = \sin{\sigma_{\text{J}}}$, with
\begin{equation}
    \sigma_{\text{J}}(\vec{B}) = \left( \sum_i \frac{\lVert \vec{J}_i \times \vec{B}_{i} \rVert}{\lVert \vec{B}_{i} \rVert} \right) / \sum_i \lVert \vec{J}_i \rVert \,.
\end{equation}
To estimate the difference from the boundary condition, we compute the deviation above the error map $\vec{B}_{error}$ as
\begin{equation}
    \vec{B}_{diff,clipped} = \max \{\vec{B}_{diff,0} - \vec{B}_{error}, 0\},
\end{equation}
where $\vec{B}_{diff,0}$ refers to the absolute difference between the modeled and observed magnetogram.
From this expression, we compute the vector norm of the clipped difference vector to quantify the deviation from the boundary condition
\begin{equation}
\label{eq:l_b0_err}
    \Delta B = \lVert \vec{B}_{diff,clipped} \rVert.
\end{equation}

\subsection{Squashing factor and twist number}

We use the squashing factor (Q-factor) $Q(x,y,z)$ and twist ($\mathcal{T}_w$) to characterize the topology of the magnetic field \citep{Titov2002, Titov2007}. 

The Q-factor is derived from the differential mapping of the magnetic field lines. 
For a given magnetic field line connecting two surfaces ($S_1$,$S_2$) with on-surface coordinates ([$x_1, y_1$], [$x_2, y_2$]), the  Jacobian matrix of differential mapping is defined as:
\begin{equation}
\underset{1\,2}{D}=\left(
\begin{array}{cc}
\frac{\partial x_2}{\partial x_1} & \frac{\partial x_2}{\partial y_1} \\
\frac{\partial y_2}{\partial x_1} & \frac{\partial y_2}{\partial y_1} \\
\end{array}
\right)
\equiv
\left(
\begin{array}{cc}
a & b \\
c & d \\
\end{array}
\right)
\label{eq:D+-}
\end{equation}
and the squashing factor (Q-factor) at $[x_1,y_1]$ can be expressed as:
\begin{equation}
Q\left(x_1,y_1\right)=\frac{a^2+b^2+c^2+d^2}
{\left|\text{det}\, \underset{1\,2}{D}\right|}. 
\label{eq:Q+-}
\end{equation}
The Q-factor is an invariant along a magnetic field line. A large Q-factor indicates the diverging nature of the local magnetic field.

The twist number ($\mathcal{T}_w$) characterizes the amount of winding along a magnetic field line, expressed as:
\begin{equation}
\mathcal{T}_w=  \int_L^{}\frac{\nabla\times\vec{B}\cdot\vec{B}}{4\pi B^2}\textrm{d}l,
\label{eq:tw}
\end{equation}
where the integral range $L$ is a segment of a magnetic field line.

In this study, we use FastQSL \citep{zhang2022fastqsl}, which utilizes graphics processing units (GPUs) to perform efficient computations of $Q$ and $\mathcal{T}_w$ for the full 3D volume.

\section{Results}
\label{sec:results}

\begin{figure*}
    \centering
    \includegraphics[width=\linewidth]{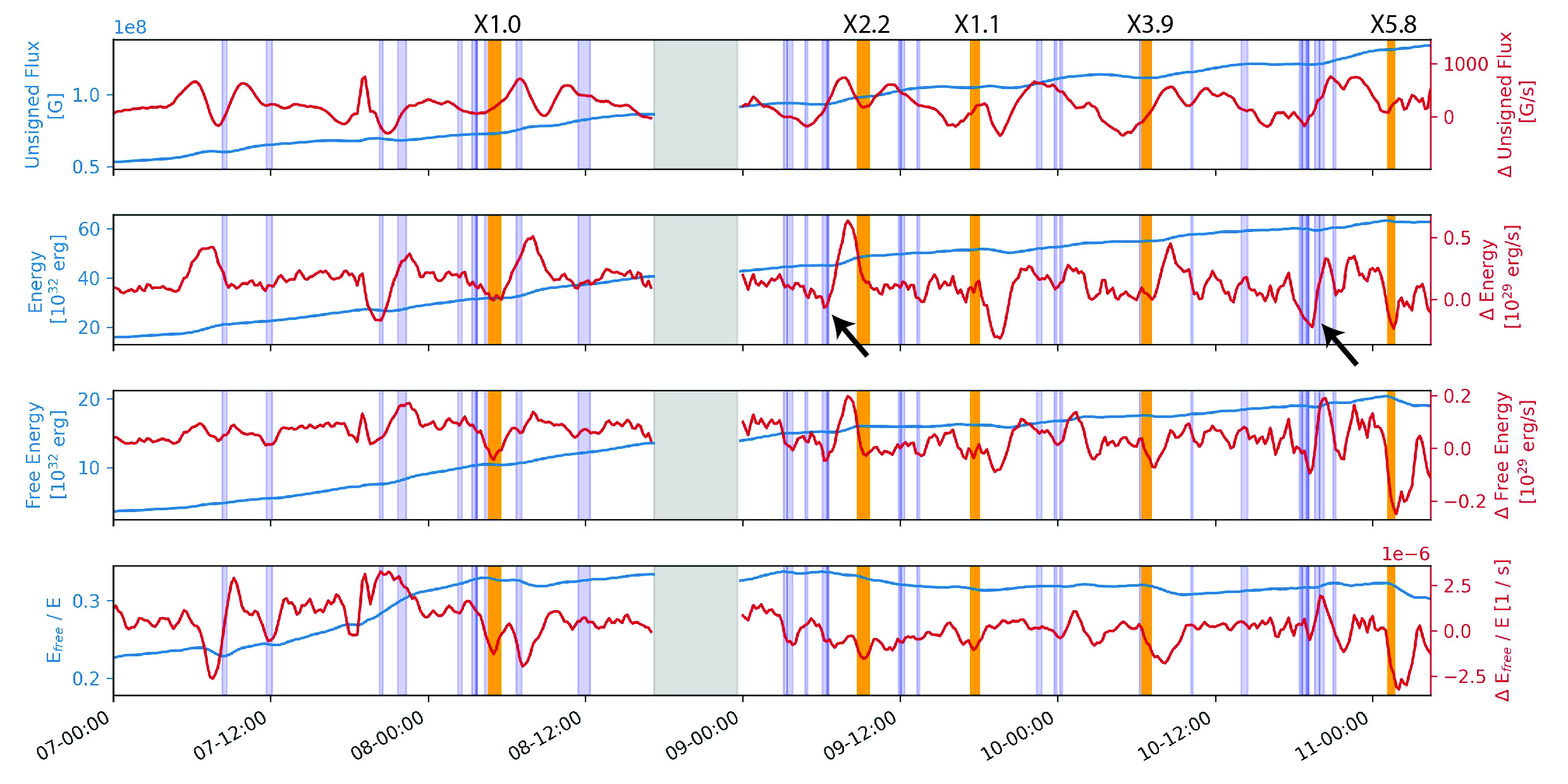}
    \caption{Overview of the temporal
    evolution of unsigned magnetic flux, magnetic energy, free magnetic energy, and the ratio between free magnetic energy and magnetic energy 
    ($E_{\text{free}}/E$) 
    of AR 13664. Blue lines correspond to the 
    direct quantities, while orange lines refer to the first order time derivatives of those quantities. Blue and orange shaded 
    bars indicate M- and X- class flares, respectively. The bars 
    outline flare start to end times, according to the GOES flare catalog. }
    \label{fig:integrated_energy}
\end{figure*}

We apply our extrapolation method to the full series of SDO/HMI SHARP vector magnetograms from 2024 May 5-00:00 to 11-04:36 UT, with a 12~min cadence. We make use of the series training approach, using the extrapolation result from the previous time step as initial condition for the training with the next boundary condition. With this approach, we only need to perform a single extrapolation from scratch, while the remaining extrapolations of the series can be obtained in real-time  \citep[about 6 minutes per extrapolation with 4 NVIDIA A100 GPUs; for details see][]{jarolim2023nf2}. Due to the HMI data gap from May 8-16:36 to May 9-00:00 UT, we separate the series into two parts, where we continue the series extrapolation from an additional initial extrapolation on May 9-00:00 UT. From our extrapolation results we render the cubes of magnetic field at a spatial sampling of 0.72 Mm per pixel. Note that the model is trained with a resolution of 0.36 Mm per pixel, however we found the reduced sampling sufficient and the decrease in data volume allows for easier access. The full resolution cubes are available through the model snapshots.

In Fig. \ref{fig:overview} we provide an overview of the ARs evolution. The GOES series in panel b shows the flare related increases in X-ray intensity. Panel c shows the radial magnetic field maps, SDO/AIA~131~$\AA$ EUV filtergrams and integrated current density throughout the modeled time sequence (dashed lines in panel b). The magnetic energy release related to flares can be seen from the change in free magnetic energy (panel a). The major flux emergence starts on 2024 May 7, as can be seen from the build-up in free magnetic energy and the magnetic field maps (red circle). In Sect. \ref{sec:results_flares} we further discuss the relation of energy depletion and the observed intensity increases in EUV. The white circles in panel c indicates supposed separatrix layers that are related to re-configurations during the solar flares (see Sect. \ref{sec:results_separatrix}).

\subsection{Magnetic energy evolution}
\label{sec:results_energy}

We compute the integrated quantities of magnetic energy $E$ (Eq.\ \ref{eq:energy}), free magnetic energy $E_\text{free}$ (Eq.\ \ref{eq:free_energy}), and the ratio of free magnetic energy to magnetic energy $E_\text{free}/E$. In addition, we compute the total unsigned magnetic flux from the SHARP vector magnetograms.

Figure \ref{fig:integrated_energy} shows an overview of the computed time series (blue lines) and the corresponding time derivatives (orange lines). The unsigned magnetic flux shows a continuous increase over the modeled time frame, which is also reflected by the energy quantities. A notable short-term flare-related change in the energy ratio only occurs for the X4.0 and X5.8 flares. This is most probably attributed to the continuous flux emergence which dominates the global energy evolution.

From the corresponding time derivatives plotted in Fig.~\ref{fig:integrated_energy}, we can identify a much better correspondence to the AR's flaring activity. 
We indicate M- and X-class flares as shaded areas from flare start- to end-time. For all the X-class flares, one can see a distinct depletion of magnetic energy, most prominently for the free magnetic energy and the ratio of free to total magnetic energy $E_\text{free}/E$. Note that the trend in the energy change profiles is still positive due to the continuous strong flux emergence. For the X1.1 flare on May 9 17:23~UT we note that the primary drop in magnetic energy occurs after the GOES flare end time. From the SDO/AIA EUV images in \textbf{Movie 1}, an increased emission at the central flare location and across the AR can be observed up to two hours past the designated flare end time, in agreement with the modeled energy decrease.
Similar to the magnetic energy release during X-class flares, also a series of M-class flares can result in a strong energy decrease, as can be seen from the flares on May 10 18:30 - 19:00 UT and on May 9 06:00 UT (black arrows). Also the M-class flares prior to the X1.0 flare on 2024 May 8 led to a noticeable drop in $\Delta E_\text{total}$.

Despite the strong energy depletion derived on May 11 01:00 UT, which is in agreement with the observed X5.8 flare, we caution that the used magnetograms are obtained close to the limb and suffer from increased uncertainties (e.g., spectropolarimetric inversions, azimuth disambiguation) and projection effects.

\subsection{Major solar flares}
\label{sec:results_flares}

\begin{figure}
    \centering
    \includegraphics[width=\linewidth]{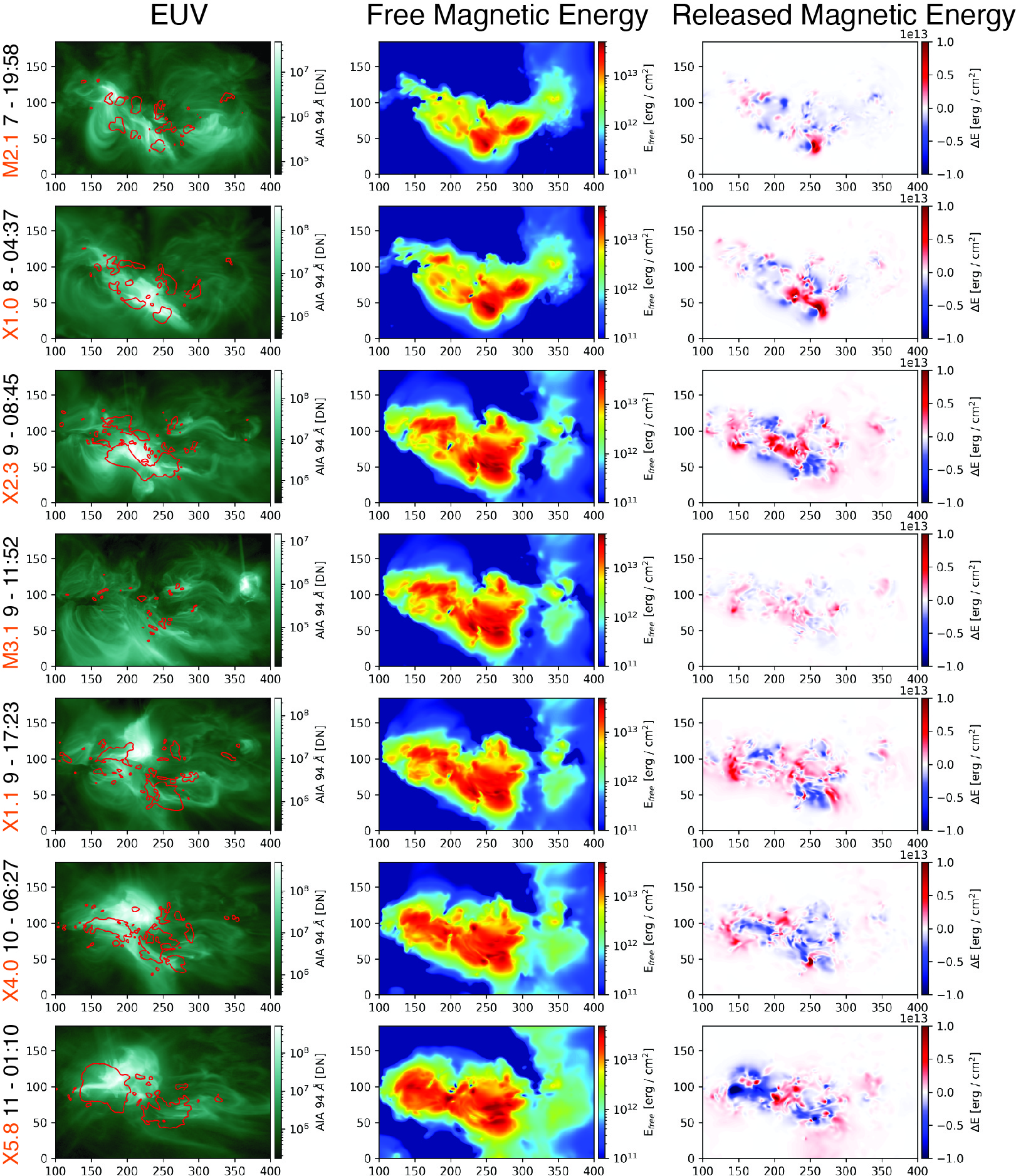}
    \caption{Major solar flares during 2024 May 7--11 and their magnetic energy distribution. The left column shows for each event the AIA 94~{\AA} EUV map time-integrated over the flare duration. The middle column shows the corresponding maps of free magnetic energy 
    $E_{\text{free}}$,
    which outline the primary regions where energy can be released. The right column shows the change in magnetic energy, indicating where energy is decreased (blue) or increased (red) over the event duration. The red contours in the integrated EUV maps indicate the 10$^{12}$ erg/cm$^{2}$ level of energy depletion ($\Delta E_{-}$). For all flare events we note a correspondence between the observed EUV emission and regions of magnetic energy change, both in terms of energy depletion (extended blue regions) and energy increase (red regions).}
    \label{fig:flares}
\end{figure}

We further investigate the change of magnetic energy during seven flare events in the spatially resolved maps, in order to identify the locations of the stored and the released free magnetic energy within the AR. We primarily consider five X-class flares that occurred between May 8 and May 11, and analyze in addition two M-class flares for context. Note that the X1.0 flare on May 8 01:33 is not related to AR 13664, the X1.0 flare on May 8 21:08~UT occurred during the HMI data gap, and the X1.5 flare on May 11 11:15~UT is too close to the solar limb to obtain reliable extrapolations.
In Fig.~\ref{fig:flares} we show maps of EUV emission, time-integrated over the duration of the event, the free magnetic energy prior to the flare eruption, and difference maps in magnetic energy between the post- and pre-flare field. The red contours outline the $10^{12}$erg/cm$^2$ threshold of the depleted free magnetic energy $E_{-}$ (Eq.\ \ref{eq:E_negative}).

The maps of free magnetic energy highlight regions with increased flaring potential. The increase in free magnetic energy directly links to the major flux emergence in the eastern part of AR 13664, and also relates to the locations of flare occurrence.

For all X-class flares we note a good agreement between locations of magnetic energy change and increased EUV emission (see Fig.~\ref{fig:flares}). The X1.0 (May 8 - 04:37 UT), the X2.3 (May 9 - 08:45 UT) and the X5.8 (May 11 - 01:10 UT) flares show additional signatures of flux emergence (red regions) during the major energy decrease (surrounding blue regions). 


For comparison, we show the same maps also for two of the more than 50 M-class flares that the AR has produced (first and fourth row in Fig.~\ref{fig:flares}).
Here, we note a much smaller energy release. Specifically, when comparing the central part of the AR during the M3.1 flare (May 9 - 11:52~UT), which occurs in the western part of the active region, the magnetic energy change mainly shows increases due to emerging flux. In contrast, for the weaker M2.1 flare (May 7 - 20:18~UT) we note regions of energy decrease that align well with the enhanced EUV emission in the central part of the active region.

In Table \ref{table:flare_energy}, we summarize the total energy difference and the released magnetic energy for the individual X- and M-class flares. Here, we compare the total energy difference ($\Delta E_\text{total}$; Eq.~\ref{eq:E_total}) and the estimated total depleted magnetic energy ($\Delta E_{-} $; Eq.~\ref{eq:E_negative}). We use the first extrapolation prior to the flare start and past the flare end, as defined by the GOES catalog, to compute the energy difference. The continuous energy build-up largely dominates the global energy evolution during the flares, which only results in an energy decrease during the two largest flares (X4.0 and X5.8) and the early M2.1 flare. When only considering the depleted magnetic energy $\Delta E_{-}$, we can provide an upper estimate of the released magnetic energy during the flare. Here, most of the X-class flares reach $>$10$^{32}$ erg, in agreement with estimates from X-ray observations, where the estimated total flare energy ranges from  $2.4 \cdot 10^{32}$ to $6.0 \cdot 10^{32}$ erg for flares between X1.0 and X2.8  \citep{woods2006flare_energy}.  The largest flare in the sequence (X5.8) reaches an upper estimate of $2.1 \cdot 10^{32}$ erg in depleted magnetic energy. The modeled M3.1 flare is one magnitude lower in energy.

\begin{table}[!h]
\centering
\caption{Energy change during the major solar eruptions. $\Delta E_\text{total}$ corresponds to the magnetic energy difference between the flare start- and end-time. $\Delta E_{-}$ refers to the integrated negative magnetic energy change over the flare duration, which provides an upper estimate of released magnetic energy during the event. The X1.0 flare on May 8 - 21:08~UT falls into the HMI data gap, and the active region is too close to the solar limb on May 11 - 11:15~UT (X1.5 flare). Therefore, both flares are not considered in this analysis.}
\label{table:height_muram}      
\centering                          
\begin{tabular}{| l l l l || r r |}        
\hline
 Date - Start & Peak & End & Class & $\Delta E_{\text{total}}$  & $\Delta E_{-} $\\ 
 $[$May 2024$]$ & & & & $[10^{31}$ erg]& $[10^{31}$ erg]\\ 
\hline                        
  \hline
  \hphantom{1}8 - 04:37 & 05:09 & 05:32 & X1.0 & 1.61 & $-8.61$ \\
  \hline
  \hphantom{1}8 - 21:08 & 21:40 & 23:10 & X1.0 & - & - \\
  \hline
  \hphantom{1}9 - 08:45 & 09:13 & 09:36 & X2.2 & 4.08 & $-12.16$ \\

  \hline
  \hphantom{1}9 - 17:23 & 17:44 & 18:00 & X1.1 & 6.17 & $-10.04$ \\
  \hline
  10 - 06:27 & 06:54 & 07:06 & X3.9 &  $-0.73$ & $-11.93$ \\
  \hline
  11 - 01:10  & 01:23 & 01:39 & X5.8 & $-5.09$ & $-21.01$ \\
  \hline
  11 - 11:15  & 11:44 & 12:05 & X1.5 & - & - \\
  \hline
\hline
  \hphantom{1}7 - 19:58 & 20:22 & 20:34 & M2.1 & $-4.40$ & $-8.64$ \\
  \hline
  \hphantom{1}9 - 11:52 & 11:56 & 12:02 & M3.1 & 2.93 & $-3.48$ \\

\hline                                   
\end{tabular}
\label{table:flare_energy}
\end{table}

\subsection{Separatrix layers}
\label{sec:results_separatrix}

\begin{figure}
    \centering
    \includegraphics[width=\linewidth]{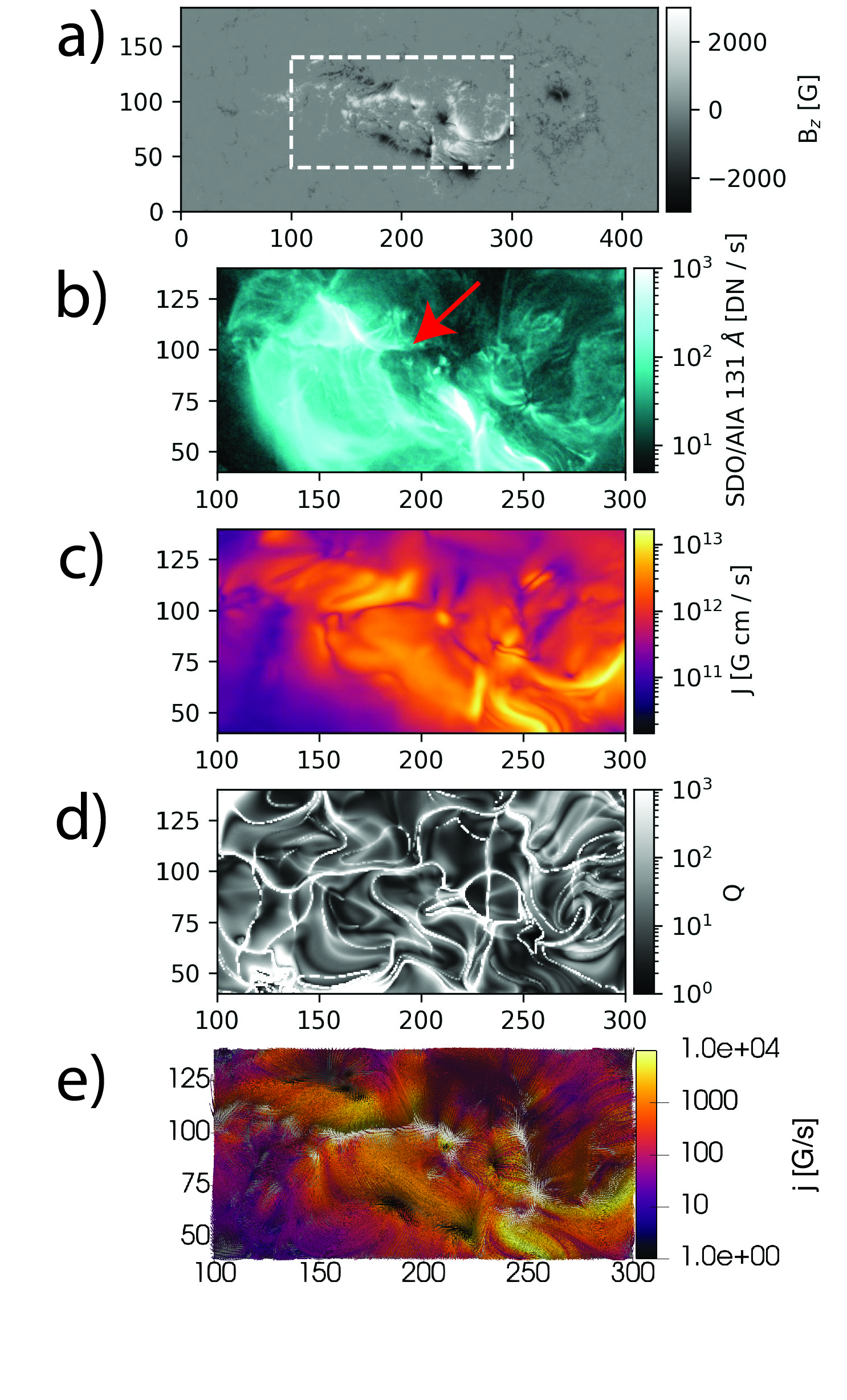}
    \caption{Example of a supposed separatrix layer and relation to the modeled magnetic field on 2024 May 8-03:00~UT. a) The observed vertical component of the magnetic field with outline of the subregion which is illustrated in panels b)--e). The red arrow indicates the supposed separatrix layer in the SDO/AIA EUV observation in the 131~{\AA} filter during a solar flare event (b). We can identify the separatrix layer as current-free layer in the vertically integrated current density (c). The squashing factor at a height of 5 Mm outlines different magnetic domains (d). Here, the separatrix layer appears as a clear separation line. From the magnetic field line plot we can further identify  the different magnetic domains in the solar atmosphere (e). The color coding refers to the local current density $j$.}
    \label{fig:separatrix}
\end{figure}

From the NLFF coronal field extrapolations, we can identify distinct layers that separate different magnetic domains \citep[i.e., quasi-separatrix layers;][]{demoulin1996qsl}. Figure \ref{fig:separatrix} shows AR 13664 during flaring activity at 2024 May 8 03:00~UT. As can be seen from the EUV maps, the flare loops clearly outline a dark separating layer (see arrow in Fig.~\ref{fig:separatrix}). We compare this observation to our modeled integrated current density, the squashing factor at a horizontal slice at 5~Mm, and a field line plot. The squashing factor indicates the cross-section of the modeled magnetic domains, where white lines represent the separation layers where field lines strongly diverge \citep{Titov2002}. Similarly, the field line plot shows the magnetic topology of the subframe of the AR. Specifically, the field line traces show the strong separation between the northern flux rope channel and the southern region. The central part of the AR ($x\approx 250$ Mm, $y\approx 75$ Mm) shows highly convoluted domains and multiple highly twisted fields.

The integrated current density shows the strongest correspondence to the EUV observation, where we can clearly identify the bright regions as twisted fields in the current density map. In particular, current-free layers show a good agreement both with lines of high squashing factor and the EUV observations. We associate these layers of low current density as a trace of quasi separatrix layers that separate magnetic domains in the active region \citep[c.f. flux-free regions; ][]{Janvier2015}. 

This is particularly relevant for the modeled flares, where we throughout observe a re-configuration in the current maps associated to solar flares. In \textbf{Movie 1} we show the temporal evolution of the current density map, where we note rapid motion and closing in of current-free layers prior to major solar eruptions. We further discuss this in Sect.~\ref{sec:X4}, where we provide an analysis of the current density evolution of the X4 flare on 2024 May 10. Figure \ref{fig:overview}c further highlights the features in the current density maps that we associate with flaring activity and the accompanying movie demonstrates the relation to the flare occurrence.

\subsection{X4.0 flare on 10 May 2024 and associated 
filament eruption}

\label{sec:X4}

\begin{figure}
    \centering
    \includegraphics[width=7cm]{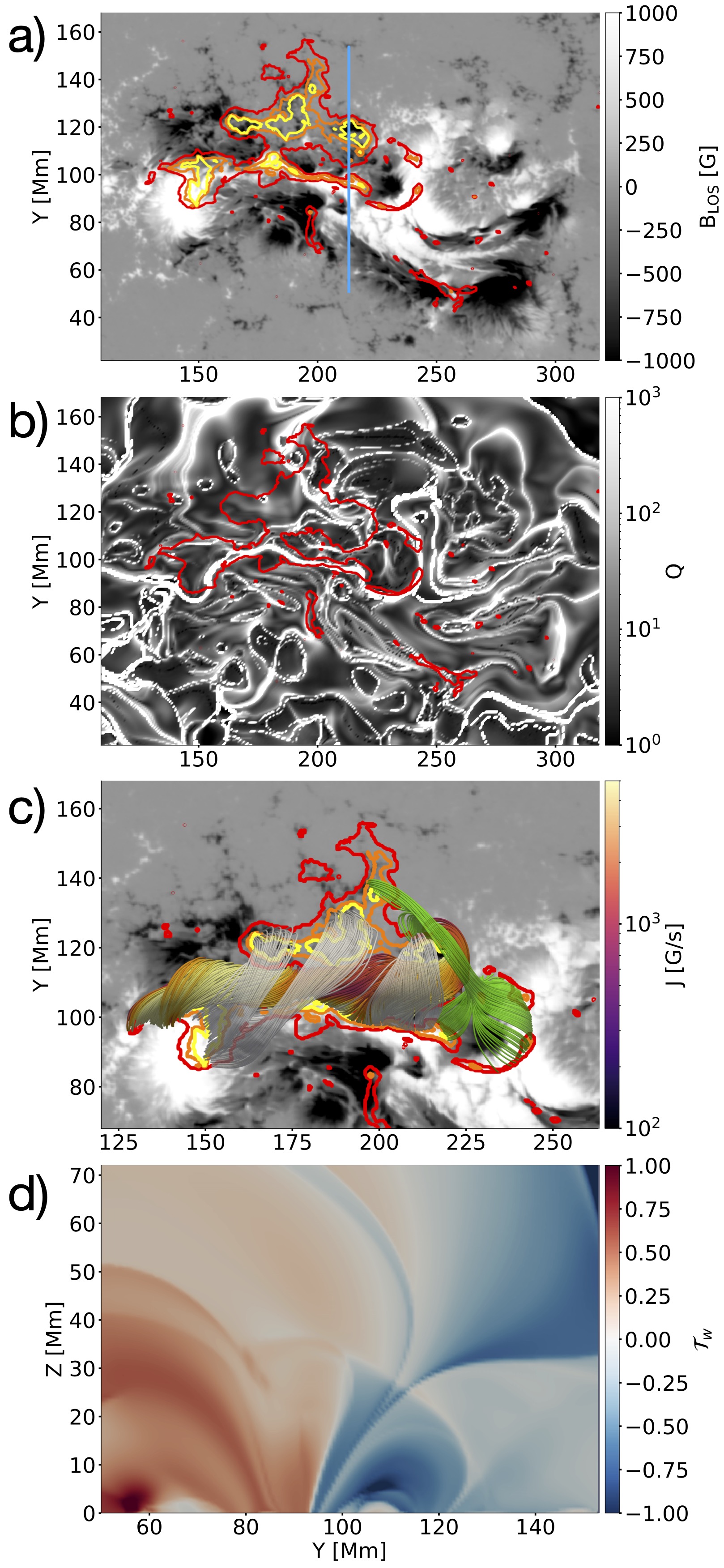}
    \caption{
    Magnetic field configuration and parameters prior to the X4.0 solar flare and associated filament eruption, as derived from a pre-flare NLFF coronal magnetic field extrapolation on 2024 May 10-05:58:43~UT.
    a) Contours of the AIA 1600~{\AA} flare ribbons on top of the pre-flare SHARP LOS magnetogram. The flare contours mark regions that exceed 150, 600, and 2400~ct/s between 06:30 and 07:00~UT. The blue line marks the position of the slice used for panel d. Panel b and c show a zoom-in of the primary flaring region.
    b) Squashing factor ($Q$) map at the bottom layer of the extrapolation volume ($z = 0 $~Mm). The red contours are the same as in panel a.
    c) Overview of selected magnetic field structures, showing the filament channel colored by the local current density, a fan-spine structure in green, and overlying fields connecting both ribbons in gray.
    d) Twist ($\mathcal{T}_w$) map along the vertical plane indicated by the blue line in panel a.
    An animation of the magnetic field line plot is available in the online journal. The series shows the evolution of the magnetic topology from 05:58:43 to 12:58:43~UT, where we note a reconfiguration of the fan-spine structure and a contraction of the central flux rope.}
    \label{fig:flare_overview}
\end{figure}

In this section, we focus on the X4.0 flare associated with a filament eruption that occurred on 2024 May 10 with a GOES start time of 06:27:00~UT and a peak time of 06:54:00~UT. Figure~\ref{fig:flare_overview} shows an overview of the pre-flare NLFF magnetic field extrapolation at 05:58:43~UT. 

Panel a shows the cumulative flare ribbons and kernels over the impulsive phase of the flare as observed by the AIA 1600~{\AA} channel. The two main flare ribbons are located in the northeastern part of the AR, with the southern ribbon extending along a narrow region of positive magnetic polarity and the northern ribbon covering a weaker region of negative magnetic polarity. The squashing factor map (panel b) shows that this main flare region was strongly separated from the rest of the AR by a separatrix layer running along the narrow positive polarity region.

\begin{figure}
    \centering
    \includegraphics[width=\linewidth]{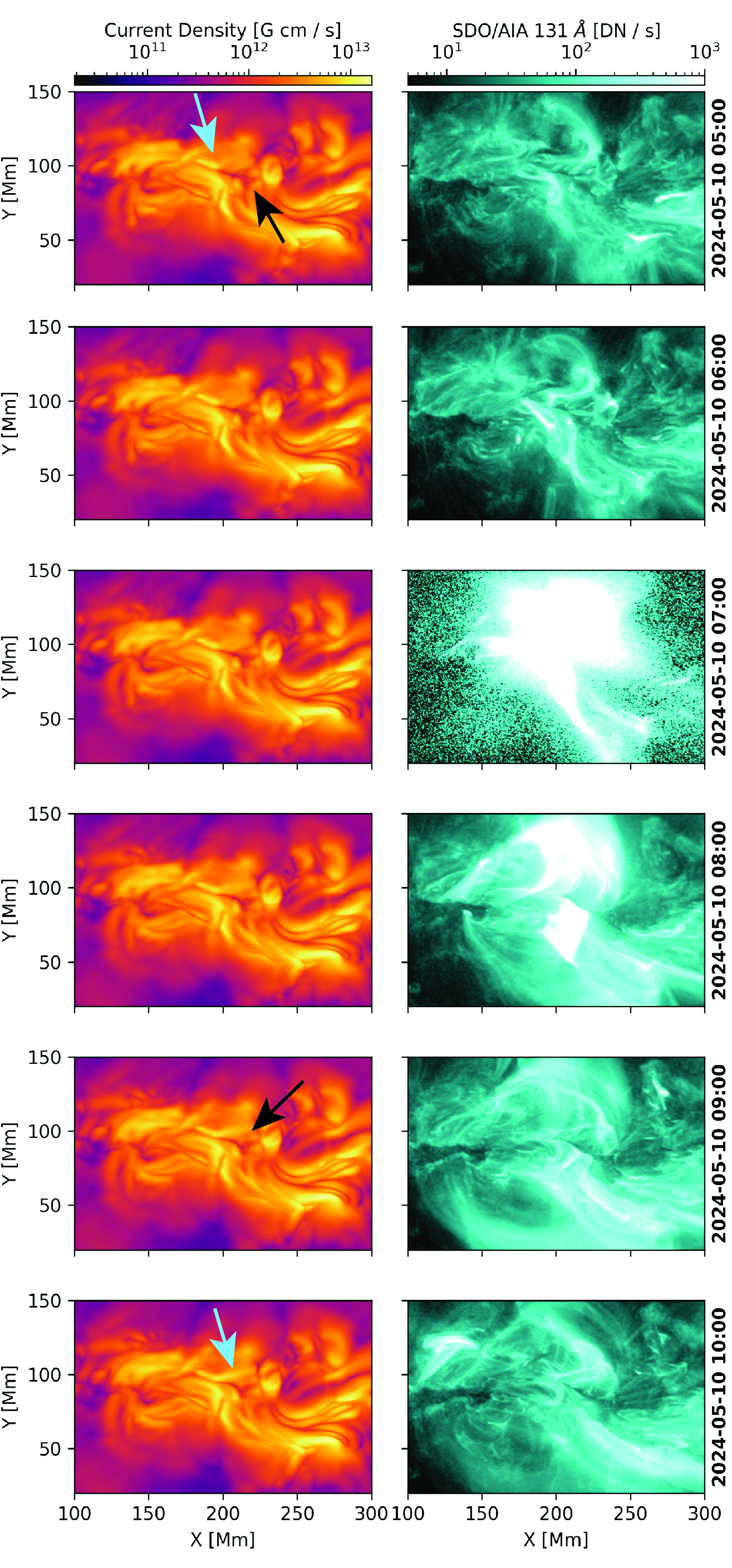}
    \caption{Comparison of the temporal evolution of the modeled vertically integrated current density (left) and the observed SDO/AIA 131~{\AA} filtergrams (right) for the X4.0 flare and filament eruption on 10 May 2024. The blue arrows indicate the reconfiguration of the magnetic flux rope, visible in the current density map. The black arrow highlights the current-free layer which undergoes substantial reconfiguration during the flare and associated filament eruption.
    }
    \label{fig:flare_sequence}
\end{figure}

This configuration resembles the one we already found two days earlier (2024 May 8), shown in Fig. \ref{fig:separatrix}. Similarly to that previous event, during the X4.0 flare the southern part of the AR is also involved in the flare, although the main eruption takes place in the strongly separated northeastern part (see the AIA 131~{\AA} time series in Fig.~\ref{fig:flare_sequence}). We observe the formation of two separate flare arcades in EUV that both connect to the southern flare ribbon in the narrow positive polarity region.

Panel c in Fig.~\ref{fig:flare_overview} shows the magnetic field configuration of the pre-flare NLFFF extrapolation, but only focuses on structures associated with the main, northern part of the event. The NLFFF extrapolation reveals strongly sheared field lines that form the filament channel and run along the polarity inversion line between the two flare ribbons.
Gray field lines start from seed sources placed in selected regions within the northern flare ribbon contours derived from AIA 1600~{\AA} maps. They reveal parts of the field structure above the filament that will be stretched by the erupting filament and reconnected beneath it, resulting in the flare ribbons and flare loops spanning between them.
Green field lines show a fan-spine structure near the western anchor point of the filament. This structure corresponds to a circular portion in the AIA 1600~{\AA} contours that connects the two flare ribbons in the west and follows the same strong separatrix layer shown in Fig. \ref{fig:flare_overview}b \citep[c.f., ][]{masson2009circular}. Some of the green field lines of the fan-spine structure closely follow the filament channel towards its western anchor point, while others extend further towards the northern part of the northern flare ribbons. This suggests that the fan-spine structure was involved in the reconnection process during the filament eruption.

The movie accompanying Fig.~\ref{fig:flare_overview} shows the temporal evolution of field structures started from fixed seed sources for each magnetic field extrapolation between 05:58:43 and 12:58:43~UT on 2024 May 10 (\textbf{Movie 2}).
The movie shows a shift of the overlying fields, a northward drift and contraction of the western half of the filament channel, and a decay of the extended connection of the fan-spine structure to the northern flare ribbon.
Note that the SHARP region moves slightly in relation to the AR during this period and the magnetic field evolves, but the seed sources from which the field lines are drawn are stationary. This means that the movie does not show the evolution of fixed field lines, and some changes can also be due to the relative motion of the magnetic field and the seed sources.

In panel d we show the twist in the pre-flare magnetic field along a cross-section through the western part of the filament. The filament channel is part of the negative twist region between Y = 100 to 120~Mm. This configuration shows an anemone type structure, where the small positive flux region (red) is overlaid by the magnetic flux rope structure (blue in panel d). This shows similarity to the mini-filament eruption mechanism proposed in  \cite{sterling2015filament} and could also be a triggering mechanism in this large-scale filament eruption.

Fig.~\ref{fig:flare_sequence} shows maps of integrated current density and SDO/AIA 131~{\AA} maps at 1 hour cadence. The blue arrow indicates the magnetic flux rope which undergoes a re-configuration during the flare event and is supposedly associated with the observed filament eruption. By comparing the flux rope before (2024 May 10-06:00) and after the eruption (2024 May 10- 08:00), a significant topology change can be observed, where the flux rope shows a larger extent after the flare eruption. The black arrow indicates the supposed separatrix configuration, where magnetic field lines of opposite polarity are compressed (c.f., squashing factor in Fig.~\ref{fig:flare_overview}). The re-configuration during the solar flare can be clearly identified from the observation on 2024 May 10-09:00~UT, where the magnetic domains are merged, and the northern part of the dark separation layer opens, forming the extended magnetic flux rope.

\section{Discussion}
\label{sec:discussion}


AR 13664 showed a rapid flux emergence from 2024 May 7 that lead to a series of M- and X-class flares, causing a multitude of X- and M-class flares and associated CMEs and being the source of the largest  geomagnetic storm with a peak Dst index of $-412$~nT of the last two decades. 

In this study, we provided an in-depth overview of the magnetic topology and evolution of AR 13664, while it was present on the Earth-facing solar hemisphere. We applied NLFF extrapolations to study the 3D magnetic topology and the evolution of the energy build up and flare-associated energy releases from 2024 May 5-00:00 to 11-04:36 UT. All our extrapolation results are publicly available (Section \ref{sec:dataavail}) as HDF5 and VTK files. We also provide the model save states, which are storage efficient ($\sim2$ MB per file), and codes for extracting and evaluating the data cube \footnote{Tutorial for data access: \url{https://github.com/RobertJaro/NF2/wiki/AR-13664}}.

The resulting data set can be used to (1) further study the relation of the solar flares to coronal mass ejections \citep{murray2018connecting}. (2) Analyze magnetic parameters from the 3D field prior to the flare eruptions \citep{Korsos2024preeruption, kusano2020flares}. (3) Analyze the temporal evolution of the modeled flaring events \citep[e.g.,][]{Purkhart2023multipoint_flares}. (4) Complement related observations, from both in-situ and remote sensing instruments \citep[e.g.,][]{Hayakawa2024}.

Our extrapolations show a realistic approximation of the observed flaring activity. The derived changes in free magnetic energy shows a direct relation to the solar flares, most prominently for the strong X-class flares and for the occurrence of multiple M-class flares over a short time frame (Sect. \ref{sec:results_energy}). In contrast, the integrated quantities of magnetic energy and free magnetic energy are subject to permanent energy increase which complicates the direct identification of flare related energy release processes. For the estimate of total released energy we are neglecting regions of energy increase, providing an upper estimate of modeled released magnetic energy. For the total depleted magnetic energy $E_{-}$, we estimate released magnetic energies in the order of $>10^{32}$ erg for X-class flares, while M-class flares typically show a lower energy difference (order of $10^{31}$ erg). To better understand NLFF modeled magnetic energy release processes in the presence of emerging flux, a comparison to MHD simulations could give further insights \citep[e.g.,][]{chen2023muram_emergence}.

We provided a detailed analysis of the X4.0 flare on 2024 May 10-06:27~UT (Sect. \ref{sec:X4}) to demonstrate how our NLFF magnetic field extrapolations can be applied to the interpretation of solar flares and filament eruptions. The magnetic field topology in a pre-flare extrapolation shows a high degree of agreement with AIA 1600~{\AA} flare ribbons and AIA EUV images. They reveal the main event geometry consisting of the filament channel, overlying loops, and a fan-spine structure. The squashing factor map derived from this extrapolation reveals a strong separatrix layer running along the southern flare ribbon, highlighting its additional magnetic connection to the southern part of the AR, which was also active during this flare. This is in agreement with previous studies that compared reconnection signatures with the location of quasi-separatrix layers \citep[e.g., ][]{dudik2014xflare, zhao2014qsl11158, dalmasse2015qsl, Janvier2015}. Our analysis suggest that we can use layers with strong gradients of current density as reference for separatrix layers that outline magnetic domains (Sect. \ref{sec:results_separatrix}). In Fig. \ref{fig:flare_sequence} we show the magnetic re-configuration during the X4.0 flare, where our model suggests the formation of a new current channel and re-configuration of magnetic domains. The strong current density build-up, close correspondence between the flare ribbons and the quasi-separatrix layer, and the topological reconfiguration, are strong evidence for magnetic reconnection \citep{aulanier2012flare_model, janvier2013flare, Janvier2015}.

Finally, with this study, we provide further extensions to PINN-based NLFF modeling, by introducing the vector potential for solenoidal simulations. While this assures that the fundamental physical law is satisfied, this leads to increased deviations from the (non-divergence free) boundary condition and requires additional compute resources.

\section{Data availability}
\label{sec:dataavail}
All our extrapolation results are publicly available.
\begin{itemize}
    \item Data: \url{https://app.globus.org/file-manager?origin_id=4263de78-cfdb-401e-a62b-dae3b935530a&origin_path=%2F}
    \item Code: \url{https://github.com/RobertJaro/NF2}
    \item Documentation for data usage: \url{https://github.com/RobertJaro/NF2/wiki/AR-13664}
\end{itemize}

With this study we provide FastQSL \citep{zhang2022fastqsl} in the NF2 framework \citep{jarolim2023nf2}.

The SDO HMI and AIA data is provided by JSOC (\url{http://jsoc.stanford.edu/}).

\section{Acknowledgments}

RJ was supported by the NASA Jack-Eddy Fellowship. 
This material is based upon work supported by the NSF National Center for Atmospheric Research (NCAR), which is a major facility sponsored by the U.S. National Science Foundation under Cooperative Agreement No. 1852977. We would like to acknowledge high-performance computing support from the Derecho system (doi:10.5065/qx9a-pg09) provided by NCAR, sponsored by the National Science Foundation.
This research was funded in part by the Austrian Science Fund (FWF) 10.55776/I4555 (AMV, SP).

This research has made use of SunPy \citep{sunpysoftware2020, sunpycommunity2020}, AstroPy \citep{2013A&A...558A..33A}, PyTorch \citep{pytorch2019_9015} and Paraview \citep{ahrens2005paraview}.

%

\vspace{5mm}
\facilities{SDO (HMI, AIA).}


\software{AstroPy \citep{2013A&A...558A..33A,2018AJ....156..123A},
          SunPy \citep{sunpycommunity2020, sunpysoftware2020, glogowski2019drms},
          Pytorch \citep{pytorch2019_9015},
          Paraview \citep{ahrens2005paraview},
          NF2 \citep{jarolim2023nf2}.
          }



\appendix

\section{Quality}
\label{sec:quality}

\begin{figure}
    \centering
    \includegraphics[width=0.7\linewidth]{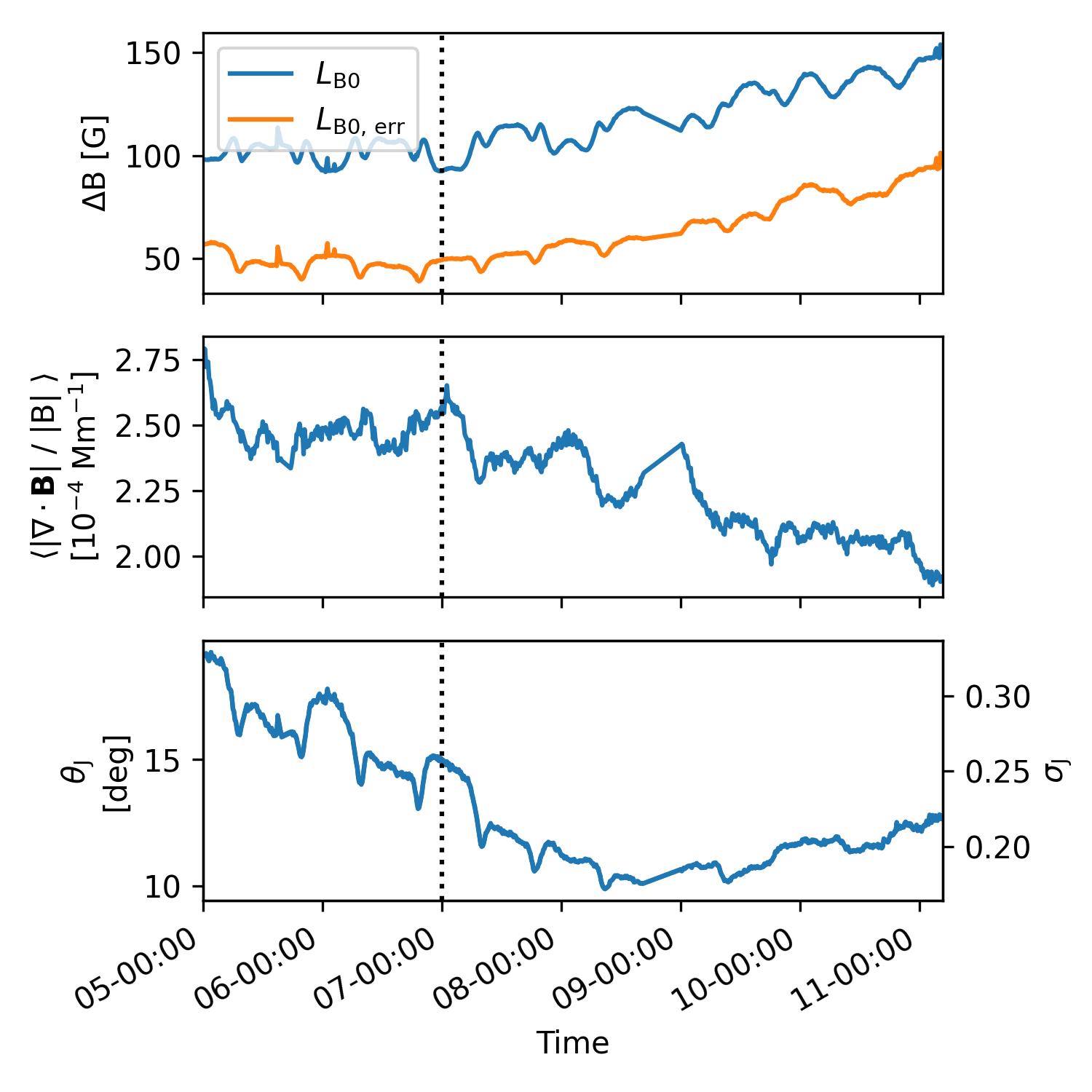}
    \caption{Metrics for divergence freeness (top) and force-freeness (bottom) over the evolution of the active region. The dashed line indicates the start of major flux emergence (c.f., Fig. \ref{fig:overview}). The decrease of $\theta_\text{J}$ can be related to the increase in magnetic currents.}
    \label{fig:quality}
\end{figure}

We compute quality metrics over the full extrapolation series, to assure the consistency of our results. In Fig. \ref{fig:quality} we compare the deviation from the boundary condition, the current-weighted angle between the magnetic field and current density ($\theta_\text{J}$), and the normalized divergence ($L_{\rm div} = |\vec{\nabla} \cdot \vec{B}| / \lVert \rm B \rVert$), as well as the arcsine of the $\theta_J$ angle. The dashed line indicates the approximate start time of increased flux emergence. Note that the divergence is computed from the grid representation based on finite-difference, while the divergence computed based on smooth derivatives is per definition zero (vector potential; $<10^{-6}$ G/Mm). As compared to the divergence estimates in \cite{jarolim2023nf2} and \cite{jarolim2024multi}, the normalized divergence is two orders of magnitude lower when the vector potential is applied. The metrics for force-freeness ($\theta_J$ and $\sigma_J$) show slightly larger values than reported in \cite{jarolim2023nf2} for the $\lambda_{ff}=0.1$ configuration. This is likely caused by the enforced divergence freeness. The decrease in $\theta_\text{J}$ over the time series is related to the flux emergence (i.e., increase in J). As can be seen from the fluctuating behaviour at the end of the sequence, the high projection angle causes significant variations. For this reason, we refrain from performing extrapolations at later points in time.

With the use of the vector potential the solutions are per definition divergence-free. However, since the observations are typically not divergence-free (due to e.g., noise, grid scaling, neglected corrugation, errors from inversion and disambiguation method) this results in an additional deviation from the boundary condition as compared to direct modeling of the magnetic field. Our uncertainty estimates further emphasize that observations close to the solar limb suffer from increased errors (see App. \ref{sec:uncertainty}).

The modeling of the vector potential has further implications for the computation requirements. The extrapolation from scratch requires about 7 hours on 4 NVIDIA A100 GPUs. For the computation of the series we require $\sim$5 min per time step with the same computational setup. Therefore, the extrapolation of the time series can still be performed in quasi real-time, but requires more computational power and computing time than direct modeling of the magnetic field B ($\approx$ 1 hour for an extrapolation from scratch).

\section{Uncertainty Quantification}
\label{sec:uncertainty}

We use ensemble modeling to estimate model dependent uncertainties. For all our model fitting we start from a randomly initialized model and use randomly sampled points during the optimization. From this, we expect that regions that are less constrained by the boundary data (e.g., weak field regions, larger errors in the magnetograms), or where the boundary condition and force-free model are in disagreement, show an increased difference among the individual extrapolations.

For our uncertainty estimation we perform five individual extrapolations and compute the standard deviation across the ensemble
\begin{equation}
    \delta B = \sqrt{\frac{\sum_i^N \left\lVert\Vec{B}_i - \overline{\Vec{B}} \right\lVert }{N}},
\end{equation}
where $\overline{\Vec{B}}$ refers to the average magnetic field vector per grid cell and $N$ to the number of ensemble runs ($N=5$).
For the visualization of uncertainty maps we compute the average uncertainty along the vertical axis.

\begin{figure*}
    \centering
    \includegraphics[width=\linewidth]{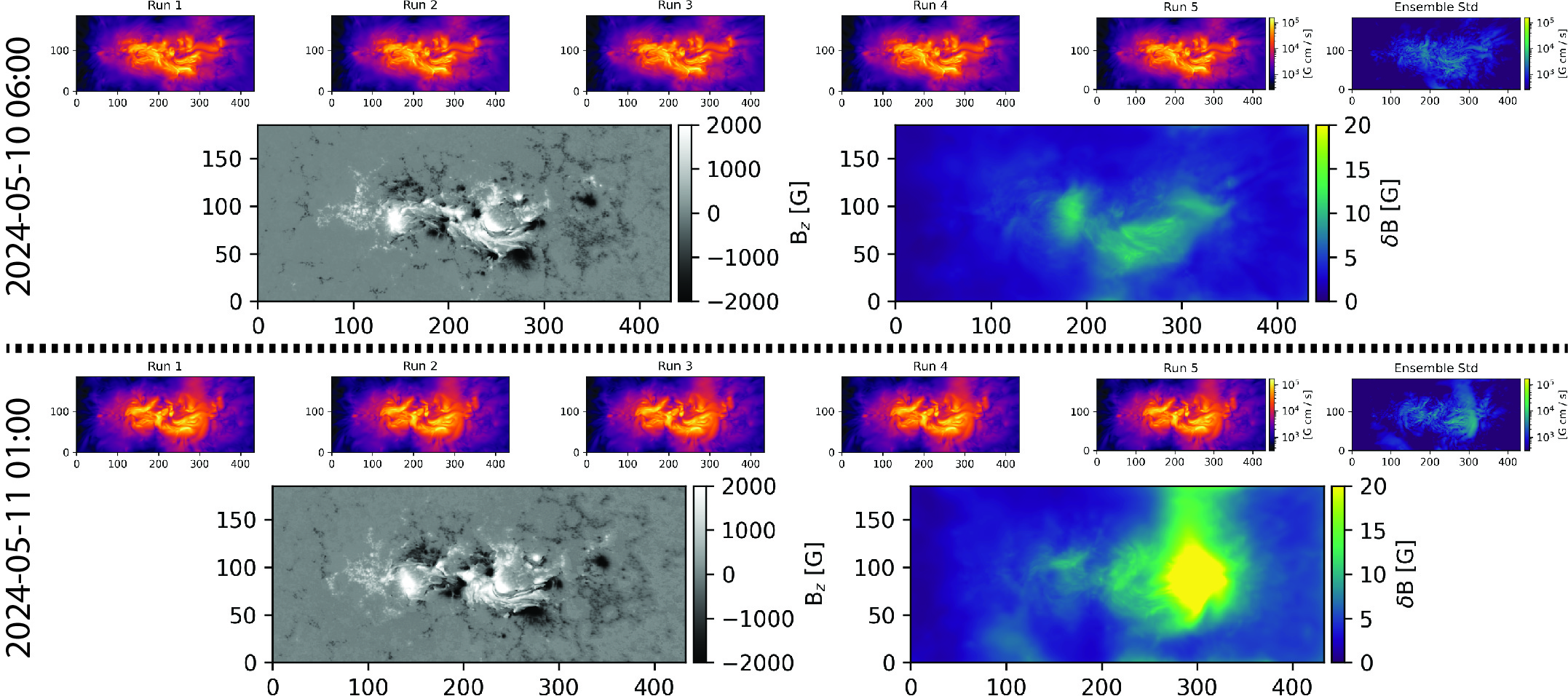}
    \caption{Uncertainty maps prior to the X4.0 and X5.8 flare. We show the vertically integrated current density of the five ensemble runs and the corresponding standard deviation at the top. The bottom plots show the reference magnetogram and the corresponding uncertainty maps. The stronger projection effects on 2024 May 11-01:00 leads to a substantial increase in uncertainty towards the west limb.}
    \label{fig:uncertainty}
\end{figure*}

The ensemble modeling requires parallel model training from scratch, therefore we can only provide uncertainty estimates for selected examples. Here, we provide uncertainty estimates for extrapolations on 2024 May 10 06:00~UT prior to the X4.0 flare, and on 2024 May 11 01:00~UT prior to the X5.8 flare. Specifically, the later extrapolation uses observations that are close to the solar limb. In Fig. \ref{fig:uncertainty} we show maps of integrated current density of the ensemble runs, the corresponding standard deviation of the current maps, and the uncertainty maps ($\delta B$). The current density maps show throughout very similar configuration, indicating that our method converges to similar solutions in terms of magnetic topology. This can also be seen from the low standard deviation.
 
The uncertainty maps highlight the regions of increased variation, which are primarily located in the southern part and the weak field region between the two polarities. Importantly, the central complex region with mixed polarities shows low uncertainties, which indicates that this region is well constrained by the boundary condition. 
For both extrapolations we note increased uncertainties close to the boundaries.
This further highlights the discrepancy of the assumed potential field boundary conditions. 
For the second extrapolation, close to the solar limb, we note a drastic increase in uncertainty, particularly towards the solar west. This suggests that extrapolations close to the solar limb should be excluded from further evaluation, particularly for points that exceed $60^{\circ}$ longitude.

\bibliography{cite}{}
\bibliographystyle{aasjournal}



\end{document}